\documentclass{sig-alternate}

\usepackage{amssymb,amsmath}
\usepackage{xcolor}
\usepackage{graphicx}
\usepackage{hyperref}
\usepackage[square, numbers]{natbib}
\usepackage{listings}
\usepackage{tabularx}

\def\doctitle{Traffic analyzer for differentiating BitTorrent handshake failures from port-scans}

\hypersetup{
  breaklinks = true,
  colorlinks = true,
  urlcolor = blue,
  citecolor = red,
  pdfkeywords = {traffic, bittorrent, tcp, port, scan},
  pdftitle = {\doctitle},
  pdfsubject = {\doctitle},
  pdfpagemode = UseNone
}
\begin{document}

\title{\doctitle}

\author{
  Kamran Riaz Khan \\
  Microsoft Corporation \\
  Redmond, WA, USA \\
  \texttt{<kakhan@microsoft.com>}
\and
  Affan A. Syed \\
  SysNet Lab, NUCES \\
  Islamabad, Pakistan \\
  \texttt{<affan.syed@nu.edu.pk>}
\and
  Syed Ali Khayam \\
  PLUMgrid Inc. \\
  Sunnyvale, CA, USA \\
  \texttt{<khayam@plumgrid.com>}
}

\maketitle

\begin{abstract}
  This paper aims to improve the accuracy of port-scan detectors by analyzing
    traffic of BitTorrent hosts and differentiating their respective BitTorrent
    connection (attempts) from port-scans.
  It is shown that by looking at BitTorrent coordination traffic and modelling
    port-scanning behavior the number of BitTorrent-related false positives can
    be reduced by $80\%$ without any loss of IDS accuracy.
\end{abstract}

\section{Introduction}

\subsection{Goal}

Many internet attacks begin by scanning a number of addresses with the goal of
  finding an open port.
This phenomenon is called ``port-scanning''.
Detecting a port-scan allows network administrators to identify an attack early
  enough to take protective measures for reducing or eliminating the ensuing
  damage.
Detection methods for port-scans account for the fact that an attacker probes a
  large number of hosts in a considerably short span of time.

Another form of traffic which exhibits behavior similar to port-scans is
  peer-to-peer communication.
In many P2P applications a host tries to initiate connections to a large number
  of \emph{peers}.
However, the presence of NATs and firewalls makes a significant number of these
  peers difficult or impossible to reach.
This results in a number of failed handshake attempts which resemble
  port-scanning attempts.
The absence of payload information in such connections also makes it impossible
  to use deep-packet inspection (DPI) for traffic labeling.

The false-positives triggered by P2P traffic in port-scan detectors makes it
  harder for security researchers and network administrators to profile network
  attacks.
Similarly the inadequacy of DPI on failed handshakes reduces the effectiveness
  of network classification tools.
To address these issues, this project inspects BitTorrent communication
  originating from a peer in order to \emph{predict} its connection attempts
  to other peers.
Consequently, those handshake failures which can be confidently labeled as P2P
  traffic are prevented from influencing results of port-scan detectors.

In addition to predicting the BitTorrent connections, this project models the
  behavior of port-scanners in order to find the port/peer ratio which can help
  differentiating between scan attempts and P2P traffic.

By using these features the goals of reducing port-scanning false-positives
  and enhancing traffic classification accuracy are achieved.
The Bro IDS is used for developing and testing the traffic analyzer and the
  statistics for improvement are based upon its default scan detector.

\subsection{Problem}

\subsubsection{Port-scanning}

Many cyber attacks need to identify potential victims before exploiting their
  weaknesses.
A popular method for discovering tractable hosts is port-scanning.
It can be defined as probing several ports on a number of machines in order
  to find susceptible hosts.
Some of the probes fail, some of them succeed but do not find any vulnerable
  service while some of them land at a port hosting a service which is further
  exploited by the attacker.

\begin{figure}[t]
  \begin{center}
  \leavevmode
    \includegraphics[width=0.4\textwidth]{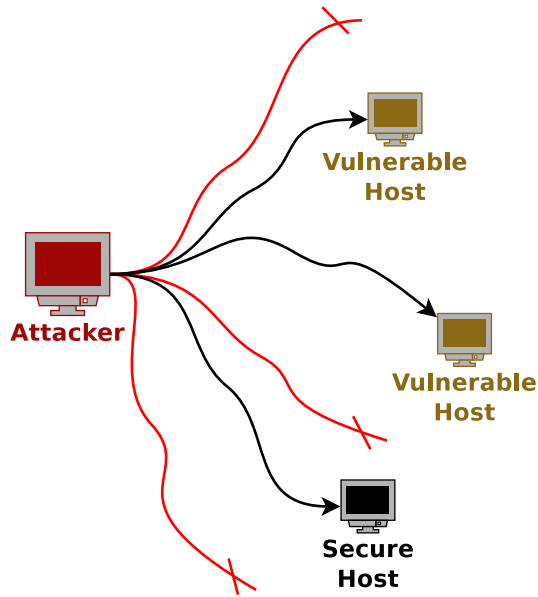}
  \end{center}
  \caption{A typical port scan}
  \label{fig:port-scan}
\end{figure}

Detecting port scans provides network administrators the ability to take
  pre-emptive measures against an attack by identifying the attacker.
Scan detectors look for $N$ events of interest across a $T$-sized time window
  \citep{Northcutt:2002}.

The first such algorithm in literature was that of Network Security Monitor
  (NSM) \citep{Heberlein:1990}.
It flagged those hosts as malicious which contacted more than fifteen
  distinct destination IP addresses within a specified time-window.

The Snort IDS \citep{Roesch:1999} uses two different pre-processors for
  scan detection.
The first is packet-oriented. It detects malformed packets such as those
  used for stealth-scanning by \emph{nmap}.
The second method is connection-oriented and is similar to the NSM algorithm.
It triggers alarm if a source IP address contacts $X$ number of ports across
  $Y$ number of IP addresses in $Z$ seconds.

The Bro IDS \citep{Paxson:1999} builds on the observation that scanners do
  not have extensive knowledge of the network topology and system
  configuration.
Therefore, it uses \emph{failed} connection attempts as indicators of
  port-scans.
For traffic using one of the services specified in a configurable list,
  Bro populates its port-scan metrics only if the connection attempts fail.
For all other traffic it keeps track of all connections regardless of their
  establishment.
When the number of distinct destination addresses crosses a particular
  threshold $N$ an alarm is triggered.

All of these methods face a number of difficulties in effective port-scan
  detection.
Temporal as well as spatial considerations impose limitations in the amount
  of connections that are tracked.
With the explosive growth of Internet many legitimate user activities
  also resemble behavior similar to port-scans.
For example, it is not uncommon for web users to initiate connections to
  dozens of different websites at once.
While such usage is catered for by Bro's strategy of tracking only
  failed connections, an issue remains in the case of P2P traffic.

P2P clients generally contact a large number of hosts in a short span
  of time and unlike other well-known services use arbitrary ports
  for their communication \citep{Karagiannis:2004:ICSP}.
This makes their behavior not only similar to port-scans but also
  harder to classify than other well-known services without the
  help of deep-packet inspection.
An example of the Bro scan detector raising false alarm is shown below:

\begin{quote}
  {\scriptsize
  \begin{verbatim}
$ bro -r bittorrent.pcap scan.bro
1317250200.756488 AddressScan 192.168.0.5 has scanned
                  20 hosts(32833/tcp)
1317250222.395888 ShutdownThresh shutdown threshold reached
                  for 192.168.0.5
1317250222.425419 AddressScan 192.168.0.5 has scanned
                  101 hosts (45612/tcp)
1317250288.527706 ScanSummary 192.168.0.5 scanned
                  a total of 326 hosts
1317250288.527706 PortScanSummary 192.168.0.5 scanned
                  a total of 224 ports
  \end{verbatim}
  }
\end{quote}

\subsubsection{P2P Handshake Failures}

\begin{figure}[t]
  \begin{center}
  \leavevmode
    \includegraphics[width=0.4\textwidth]{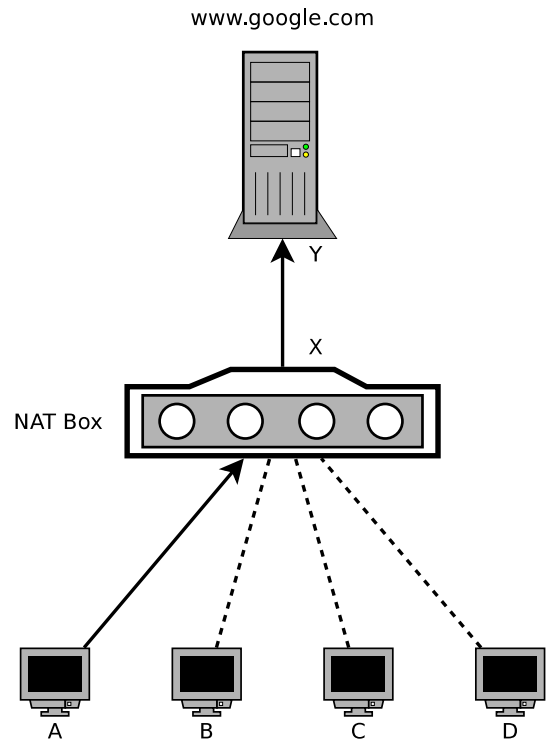}
  \end{center}
  \caption{Outbound NAT connections}
  \label{fig:nat-outbound}
\end{figure}

\begin{figure}[t]
  \begin{center}
  \leavevmode
    \includegraphics[width=0.4\textwidth]{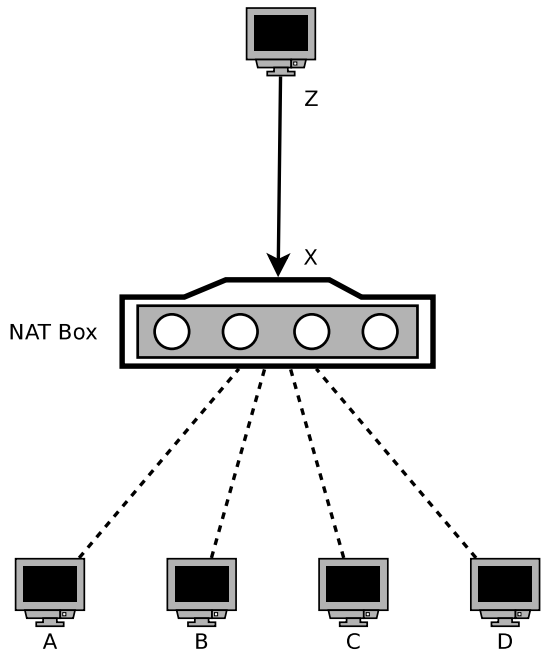}
  \end{center}
  \caption{Inbound NAT connections}
  \label{fig:nat-inbound}
\end{figure}

Network Address Translator (NAT) are devices that allow multiple hosts
  on a private network to communicate with the Internet using a single
  public IP.
This is achieved by modifying network layer information on-the-fly.
When an internal host initiates a connection to an outsider a dynamic
  mapping is created which allows the response to be forwarded to that
  particular host as shown in \autoref{fig:nat-outbound}.
Unsolicited inbound connections are dropped since NAT simply does not
  know which internal host should receive the data.
\autoref{fig:nat-inbound} shows a NAT box receiving incoming data but does not
  know which internal client $[A, D]$ should the data be forwarded to.

Firewalls are devices which inspect traffic passing through them and
  filter it based on a set of rules.
In basic implementations unsolicited inbound connections are rejected
  while outbound connections are allowed.

Hosts which are behind NAT and/or firewalls have restricted connectivity
  in terms of receiving inbound connections.
In P2P terminology they are called \emph{unconnectable peers}.
The ratio of such peers is considerably high, ranging from 60\% to 90\%
  of all P2P users \citep{Dacunto:2010,Dacunto:2009}.

When a P2P client needs to contact other peers it uses a number of
  methods to obtain a list of their addresses.
The most primitive of these methods is to contact a central tracker
  and request \emph{peer lists}.
Other de-centralized methods such as distributed hash tables also exist
  for storing peer lists.

However, in popular protocols such as BitTorrent the peer lists
  returned to a client via these methods usually contains information
  about other peers regardless of their connectivity status.
As a result, unconnectable peers are often present in these lists.

\begin{figure}[t]
  \begin{center}
  \leavevmode
    \includegraphics[width=0.4\textwidth]{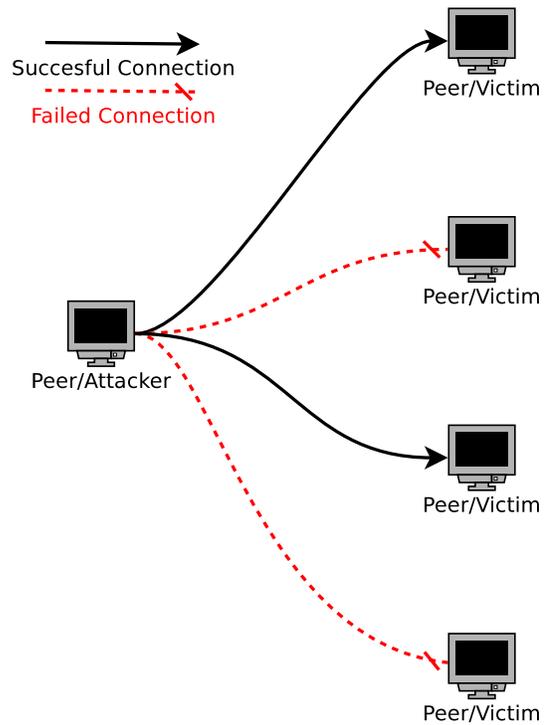}
  \end{center}
  \caption{Connections of a P2P host and a port scanner}
  \label{fig:problem-p2p-portscan}
\end{figure}

Unconnectable peers result in a large number of failed connection
  attempts which can negatively influence the false-positive rate of
  scan detectors such as Bro.
The similarity of connectivity behavior of P2P hosts and a scanner is
  shown in \autoref{fig:problem-p2p-portscan}

\subsection{Solution}
\subsubsection{BitTorrent Communication Analysis}

BitTorrent is the most popular P2P file-sharing protocol, accounting for more
  than half of all P2P file sharing traffic \citep{opaque:2009}.
Files shared via BitTorrent are divided into equal-sized chunks.
\emph{Peers} download these pieces from each other while uploading them
  simultaneously to other peers as well.
All the peers which are sharing a particular file at a time are collectively
  called a \emph{swarm}.

\begin{figure}[t]
  \begin{center}
  \leavevmode
    \includegraphics[width=0.4\textwidth]{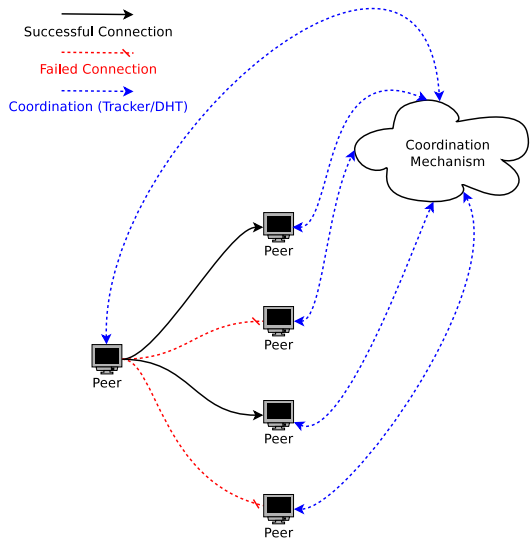}
  \end{center}
  \caption{BitTorrent coordination}
  \label{fig:solution-p2p-portscan}
\end{figure}

Before peers start to download a file however, they need to coordinate first in
  order to locate each other.
The BitTorrent coordination lines in \autoref{fig:solution-p2p-portscan} take
  place \emph{before} the actual traffic and hence can be used to predict the
  successful and failed connections originating from a peer.

The earliest method for BitTorrent coordination was via centralized
  \emph{trackers}.
Centralized trackers keep track of which peers are downloading a file.
When a new peer wants to join the swarm it requests a peer list from the tracker.
The tracker responds with a list of $k$ randomly picked peers, containing both
  connectable and unconnectable hosts.
Connection attempts to unconnectable hosts then subsequently results in
  failed TCP connections \citep{Liu:2009}.

In order to alleviate the load on trackers as well as increase resilience
  against take-down attempts two de-centralized methods exist for obtaining
  peer lists: Distributed Hash Table (DHT) and Peer Exchange (PEX).

A DHT is a distributed database overlayed over a network of computers called
  nodes.
All nodes can store $(key, value)$ pairs on the DHT and search for values via
  key-lookups.
In order to join a DHT a peer has to first contact a \emph{bootstrap} node
  in that DHT.
Currently, the BitTorrent community uses two major DHTs: the Azureus DHT and
  the Mainline DHT.
The Azureus DHT is in use only by the Azureus clients while the Mainline DHT
  is used by other clients such as $\mu$Torrent, BitComent and Mainline.
Both DHTs are implementations of the Kademlia protocol \citep{Crosby:2007}.

The second de-centralized method of peer discovery is PEX.
In PEX peers ``gossip'' with each other in regular intervals about the list
  of active peers they know about \citep{Wu:2010}.
There is no official specification for the peer exchange protocol in
  BitTorrent.
The de facto standard for Peer Exchange is $\mu$Torrent PEX.

All these methods of peer discovery can be interpreted by a traffic analyzer,
  with the exception of trackers that serve on HTTPS.
By reading the peer lists returned via HTTP tracker, DHT or PEX it is possible
  to ``predict'' the BitTorrent connection attempts that are going to originate
  from a given source IP.
This shall in turn make it possible to label failed connection attempts and
  differentiate them from port-scans.

\begin{figure}[t]
  \begin{center}
  \leavevmode
    \includegraphics[width=0.4\textwidth]{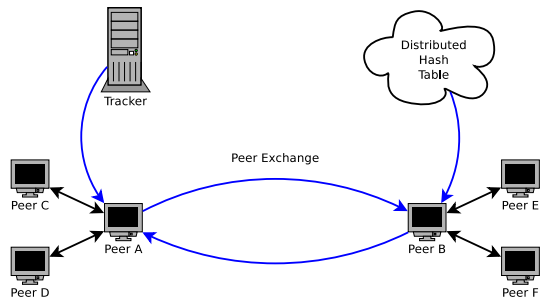}
  \end{center}
  \caption{Various methods for BitTorrent coordination}
  \label{fig:bittorrent-coordination}
\end{figure}

\section{Implementation}

\subsection{Bro IDS}

The Bro IDS was used to develop and test the traffic analyze along with its scan
  detector for providing baseline statistics.
During the course of development Bro's core was modified to provide the
  following helper functions:

\begin{itemize}
  \item \autoref{lst:bro_raw_ns_bytes_to_uint16}:
    \texttt{raw\_ns\_bytes\_to\_uint16} \\
    Interpret the first two bytes of a string as a 2-byte unsigned integer in
    network-byte order.
  \item \autoref{lst:bro_raw_nl_bytes_to_uint32}:
    \texttt{raw\_nl\_bytes\_to\_uint32} \\
    Interpret the first four bytes of a string as a 4-byte unsigned integer in
    network-byte order.
  \item \autoref{lst:bro_sub_bytes_sane}:
    \texttt{sub\_bytes\_sane} \\
    Bro's default \texttt{sub\_bytes} function had the idiosyncratic behavior in
    that it if the indices were greater than 0 it decremented them by 1 before
    slicing the string.
    This function works around this behavior by retaining uniformity of indices.
  \item \autoref{lst:bro_raw_byte}:
    \texttt{raw\_byte} \\
    Returns the raw byte at a particular index in a string as an unsigned integer.
  \item \autoref{lst:bro_raw_bytes}:
    \texttt{raw\_bytes} \\
    Returns the raw bytes in a string as a vector of unsigned integers.
\end{itemize}

\subsection{Central Trackers}

A central tracker keeps track of peers that are currently downloading a file and
  transfers this list to new peers that are trying to join the swarm.

\begin{figure}[t]
  \begin{center}
  \leavevmode
    \includegraphics[width=0.4\textwidth]{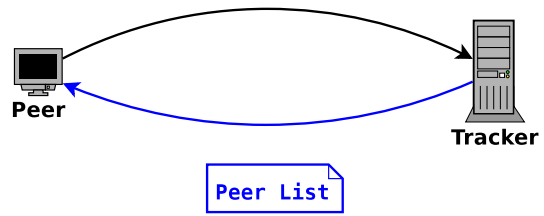}
  \end{center}
  \caption{Central tracker}
  \label{fig:central-tracker}
\end{figure}

There are two types of central tracker used by BitTorrent clients.

\subsubsection{HTTP Tracker}

A built-in analyzer was available in Bro for the HTTP tracker communication.
However, using the peer lists extracted from this analyzer did not yield any
  connection hits.
Therefore we manually analyzed the contents of TCP packets to extract peer lists
  whenever a specific regular expression is matched in the packet.
This regular expression confirms the presence of peer list in the benc format
  that used by BitTorrent.

Any peers that are found in the peer lists are added to a global
  \texttt{peer\_mappings} table which is indexed by source IPs and yields
  the (IP, Port) tuple for targets extracted from the peer list.

For example, if client $X$ gets a peer list in which a peer $A$ is listed as
  listening on port $6668$, $(A, 6668)$ is added as a possible target for $X$.

\autoref{lst:httptracker_tcp_contents} shows the code used by the HTTP analyzer
  to process each TCP packet.

\subsubsection{UDP Tracker}

The BitTorrent ecosystem is moving on to UDP trackers because of less overhead
  compared to HTTP trackers.
Many major tracker websites such as The Pirate Bay and Mininova have already
  switched to UDP trackers.
Bro did not have a built-in analyzer for the UDP tracker communication so we
  analyzed UDP traffic for packets of particular lengths and matched the
  \texttt{action} fields of the packet for \texttt{annount\_response}.
Once the UDP tracker communication was identified, \texttt{announce\_response}
  helper function is called which in turn extracts the peer list and stores it
  in a global \texttt{peer\_mapping} table.

\autoref{lst:udptracker_udp_contents} shows the processing of each UDP packet
  by the UDP tracker analyzer.

\subsection{Distributed Hash Tables}

\subsubsection{Azureus DHT}

\autoref{tab:adht-request-header} and \autoref{tab:adht-reply-header} show the
  packet structures for headers of ADHT packets.
ADHT requests start with a random connection ID and do not have a fixed length.
To identify these requests, the 16th byte of each UDP packet is checked for a
  valid ADHT protocol version.
Based on the found protocol version, various offsets are added to a running
  counter corresponding to different fields.

Eventually, the originating port of the packet should appear in network-byte
  order at a specific offset.
If that check does succeed, the packet is a valid ADHT request and bytes
  $[8, 12]$ are checked for \texttt{ACTION} field of the request.

Upon finding a valid \texttt{ACTION}, either a helper \texttt{find\_*\_request}
  function is called which in turn extract the \texttt{TRANSACTION\_ID} and
  store it in a global table.

Similarly, all packets are checked for valid \texttt{TRANSACTION\_ID}s from
  requests that have already been seen. If one is found, a helper
  \texttt{find\_*\_response} function is called which extracts the peer list
  out of the DHT values.

\autoref{lst:adht_udp_contents} shows the code used by the ADHT analyzer to
  process each UDP packet.

\subsubsection{Mainline DHT}

For extracting Mainline DHT peer lists regular expressions are matched inside
  UDP packets for ``\texttt{nodes}'' and ``\texttt{values}'' responses.
Upon matching the regular expressions the number of peers is determined and then
  each peer's IP address and port number is extracted in a sequential manner.
\autoref{lst:mdht_find_values} shows the code used by the MDHT analyzer to
  extract peers out of DHT respones.

\subsubsection{Other DHTs}

During the course of development we observed that certain UDP traffic flows
  between BitTorrent hosts before they initiate TCP connections.
Matching this UDP traffic allows us to predict the TCP connections since the
  destination port remains same on both transport layer protocols.
The signatures for matching miscellaneous DHT traffic were taken from
  \texttt{libprotoident}'s database, which is an open-source library for
  detecting application layer protocols from length of a payload combined with
  its first four bytes.

\autoref{lst:btudp_udp_contents} shows the code used by BTUDP analyzer to match
  signatures across each UDP packet.
Once a BTUDP packet is detected, the target IP and port are added to safe
  \texttt{peer\_mappings}.

\subsection{Peer Exchange}

\begin{figure}[t]
  \begin{center}
  \leavevmode
    \includegraphics[width=0.4\textwidth]{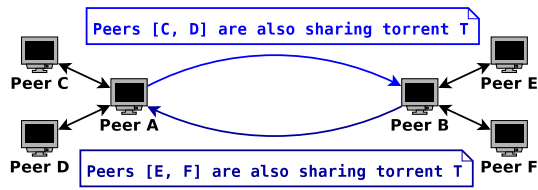}
  \end{center}
  \caption{Peer Exchange}
  \label{fig:pex}
\end{figure}

Compatible BitTorrent clients ``gossip'' with each other about the clients they
  know exist in their swarm.
This form of coordination is called Peer Exchange and is shown in
  \autoref{fig:pex}.

\subsubsection{UTPEX}

$\mu$Torrent PEX or UTPEX is the de facto standard of Peer Exchange for
  BitTorrent clients.
The packets are searched for regular expressions matching ``added'' responses
  which are sent by a client to its peers existing when it learns about new clients
  in the swarm.
\autoref{lst:utpex_tcp_contents} shows the code used by UTPEX analyzer for
  processing each TCP packet.

\subsection{Port/Peer Ratio}

Port/Peer Ratio is a feature which relies on behavioral differences between port
  scanners and BitTorrent downloaders.
Before explaining these differences it is important to outline various
  categories in which a scanner's behavior can be classified.

\begin{figure}[t]
  \begin{center}
  \leavevmode
    \includegraphics[width=0.4\textwidth]{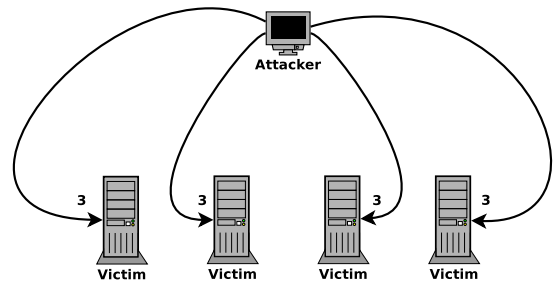}
  \end{center}
  \caption{A horizontal scan}
  \label{fig:horizontal-scan}
\end{figure}

There are three kinds of port scans \citep{Lee:2003}.
The first scan type is ``horizontal'' in which the attacker probes multiple
  hosts on the same port as shown in \autoref{fig:horizontal-scan}.
The motiviation for such a scan stems from the fact that a known vulnerability
  would exist on that port.

\begin{figure}[t]
  \begin{center}
  \leavevmode
    \includegraphics[width=0.4\textwidth]{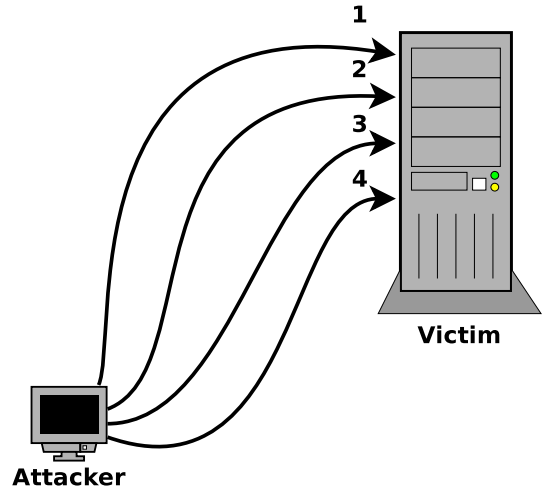}
  \end{center}
  \caption{A vertical scan}
  \label{fig:vertical-scan}
\end{figure}

The second scan type is ``vertical'' in which the attack probes a single host
  for multiple ports as shown in \autoref{fig:vertical-scan}.
The motivation for such a scan lies in attacker targetting a specific victim to
  find any vulnerability he can find.

\begin{figure}[t]
  \begin{center}
  \leavevmode
    \includegraphics[width=0.4\textwidth]{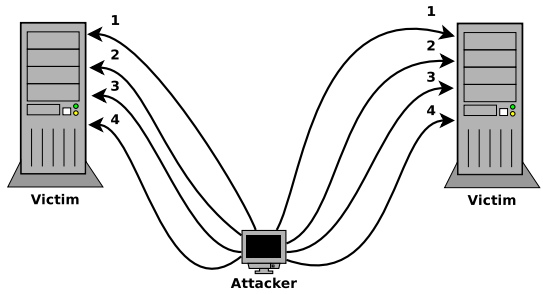}
  \end{center}
  \caption{A hybrid scan}
  \label{fig:hybrid-scan}
\end{figure}

A hybrid or block scan is a combination of both horizontal and vertical scans
  and is shown in \autoref{fig:hybrid-scan}.

\begin{figure}[t]
  \begin{center}
  \leavevmode
    \includegraphics[width=0.4\textwidth]{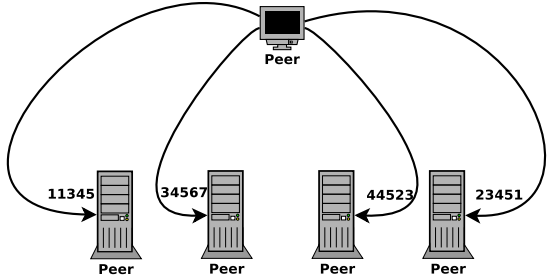}
  \end{center}
  \caption{Port connectivity behavior of a P2P client}
  \label{fig:p2p-scan}
\end{figure}

In contrast, a P2P host contacts a large number of hosts on ports that are
  fairly random and almost unique since most of the hosts are behind NAT and
  have set up port-forwarding on non-standard ports.
This is shown in \autoref{fig:p2p-scan}

To use these observations, Port/Peer Ratio $PPR$ of each host is kept track of.
Ports correspond to unique ports while peers correspond to number of unique IP
  addresses that a hose has probed.
From the discussion above:

\begin{itemize}
  \item If $PPR$ is close to $0$, it indicates a horizontal scan.
  \item If $PPR$ is close to $1$, it indicates P2P behavior.
  \item If $PPR$ is greater than $1$, it indicates a vertical or hybrid scan.
\end{itemize}

\autoref{lst:btscan_check_scan} shows code modifications to Bro's scan detector
  for suppressing alarms when $PPR$ of the attacker lies between a lower and
  upper threshold.
By default these thresholds are specified as $0.75$ and $1$ respectively.

\section{Results}

\subsection{Controlled Experiment}

For this experiment labeled data sets containing port scanners and BitTorrent
  hosts were used to gauge the IDS performance.
With support of Dr. Ali Khayam we were able to obtain traces for 41 TCP port
  scanners which were generated at various rates for the RAID '10 paper
  \citep{Haq:2010} on impact of P2P on anomaly detection.
In addition to the RAID data sets, \texttt{nmap} was used to generate traffic
  for 10 further scanners.
The combined traces had 51 port scanners which scanned at rates varying from
  $0.01/s$ to $1000/s$.

In order to process P2P traffic, BitTorrent traces were generated on 49
  different virtual hosts.
Various BitTorrent clients including Azureus, BitComet, Deluge, $\mu$Torrent,
  Transmission and FlashGet were used to generate the traffic on both Windows
  and Linux platforms.
For each client application the default settings were used which resulted in
  unencrypted traffic for all hosts.
It was ensured that the hosts were not infected at the time of traffic
  generation.

The traffic for both the scanners and BitTorrent hosts were aligned temporally
  to start at the same time while IP addresses were rewritten to avoid spatial
  collisions.
The IP addresses were reassigned according to the criterion which assigned
  BitTorrent hosts and port scanners separate networks for ease of labeling.

\begin{figure}[t]
  \begin{center}
  \leavevmode
    \includegraphics[width=0.4\textwidth]{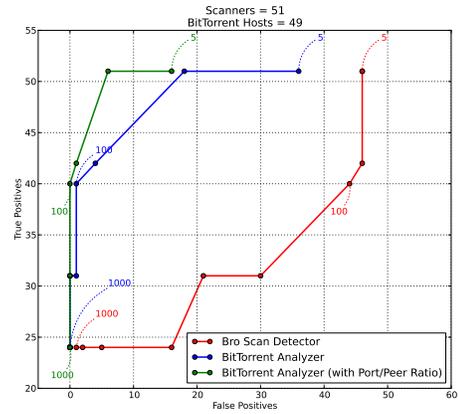}
  \end{center}
  \caption{ROC curves for port scanner performance}
  \label{fig:roc-curves}
\end{figure}

To compare the performance of scan detector and its improvements ROC curves were
  used.
The results are shown in \autoref{fig:roc-curves}.
Each point on the graph shows different thresholds for the number of addresses a
  host can probe within the 15 minute time window before being flagged as a
  scanner.
At more relaxed thresholds, e.g., $1000$, the points concentrated in the
  lower-left corner of the graph.
Similarly, at stricter thresholds, e.g., $5$, the points moved towards the
  upper-right corner of the graph as more scanners and BitTorrent hosts were
  flagged.

\subsection*{Conclusion}

The original Bro scanner had a curve which bent significantly towards the
  lower-right corner of the graph.
Adding the BitTorrent analyzer engine improved the performance of scan detector
  significantly without any loss of IDS accuracy since the curves do not intersect
  at any threshold.
At the default Bro threshold, i.e., $100$, the false positive rate is decreased
  from $\frac{44}{49} = 89\%$ to $\frac{1}{49} = 2\%$.
By using the Port/Peer Ratio feature, the false positive rate is further
  improved to $\frac{0}{49} = 0\%$.

\subsection{Live Traffic}

For live traffic experiments a $400$ GB trace was used which was captured at a
  Nayatel B-RAS over a duration of $24$ hours.

\begin{figure}[t]
  \begin{center}
  \leavevmode
    \includegraphics[width=0.4\textwidth]{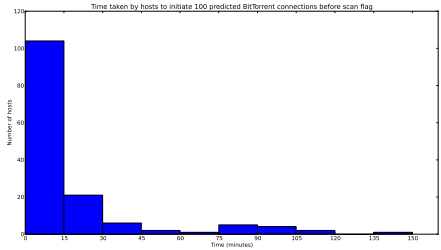}
  \end{center}
  \caption{Histogram depicting duration taken by BitTorrent hosts to initiate
           $100$ predicted connections before being flagged as a scanner}
  \label{fig:nayatel-histogram-intervals-before-flag}
\end{figure}

We first ran BitTorrent Analyzer to figure out which connections could have been
  ``predicted'' by looking at the BitTorrent coordination traffic.
Then we calculated the duration each BitTorrent host took to contact $100$
  predicted connections \emph{before} being flagged as a scanner.
So if host $A$ was flagged as a scanner at 9:05 AM and $100$ predicted
  BitTorrent connections originated from $A$ between 9:00 and 9:05 AM the
  duration would be $5$ minutes.
A histogram was generated with each bin denoting $15$ minutes of time.
As shown in \autoref{fig:nayatel-histogram-intervals-before-flag}
  most of the BitTorrent hosts generated $100$ predicted BitTorrent connections
  in the first $15$ minutes bin before being flagged as a scanner.
The results indicate a strong influence of predicted BitTorrent connections on
  the scan flags.

\begin{figure}[t]
  \begin{center}
  \leavevmode
    \includegraphics[width=0.4\textwidth]{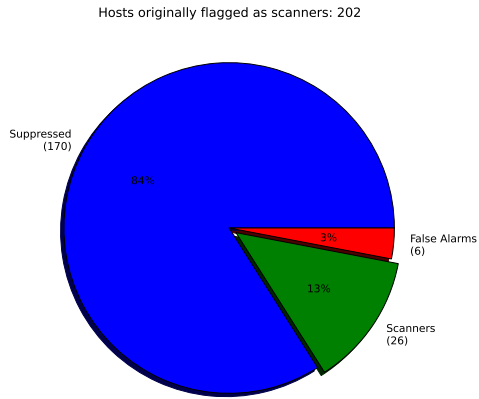}
  \end{center}
  \caption{Analysis of scan flags on the $400$ GB live traffic trace}
  \label{fig:nayatel-flagstats-pie}
\end{figure}

Bro originally detected $202$ scanners in the $400$ GB trace.
By filtering connections through the BitTorrent Analyzer the number of alarms was
  reduced to $32$ as $170$ flags were suppressed.
We manually analyzed the remaining $32$ hosts to observe that $26$ of them were
  actual scanners while $6$, or $3\%$ of the original flags, were false alarms
  that were still being raised.
These results are summarized in the pie-chart shown in
  \autoref{fig:nayatel-flagstats-pie}.

\subsection*{Conclusion}

We concluded that almost all BitTorrent hosts generate false alarms because of
  their aggressive connectivity behavior.
The BitTorrent Analyzer was able to reduce the number of false alarms
  significantly, however about $3\%$ of the total flags still generated false
  positives.

\renewcommand{\bibsection}{\section{\refname}}
\bibliographystyle{acm}
\bibliography{report}

\begin{thebibliography}{10}

\bibitem{Crosby:2007}
{\sc Crosby, S.~A., and Wallach, D.~S.}
\newblock An analysis of {B}it{T}orrent's two {K}ademlia-based {DHT}s, 2007.

\bibitem{Dacunto:2010}
{\sc D'Acunto, L., Meulpolder, M., Rahman, R., Pouwelse, J., and Sips, H.}
\newblock Modeling and analyzing the effects of firewalls and {NAT}s in {P2P}
  swarming systems.
\newblock In {\em IEEE IPDPS 2010 (HotP2P 2010)\/} (April 2010), pp.~1--8.

\bibitem{Dacunto:2009}
{\sc D'Acunto, L., Pouwelse, J., and Sips, H.}
\newblock A measurement of {NAT} \& firewall characteristics in peer-to-peer
  systems.
\newblock In {\em Proc. 15-th ASCI Conference\/} (June 2009), Advanced School
  for Computing and Imaging (ASCI), pp.~1--5.

\bibitem{Haq:2010}
{\sc Haq, I.~U., Ali, S., Khan, H., and Khayam, S.~A.}
\newblock What is the impact of {P2P} traffic on anomaly detection?
\newblock In {\em Proceedings of the 13th international conference on Recent
  advances in intrusion detection\/} (Berlin, Heidelberg, 2010), RAID'10,
  Springer-Verlag, pp.~1--17.

\bibitem{Heberlein:1990}
{\sc Heberlein, L.~T., Dias, G.~V., Levitt, K.~N., Mukherjee, B., Wood, J., and
  Wolber, D.}
\newblock A network security monitor.
\newblock {\em Security and Privacy, IEEE Symposium on\/} (1990), 296.

\bibitem{Karagiannis:2004:ICSP}
{\sc Karagiannis, T., Broido, A., Brownlee, N., Claffy, K.~C., and Faloutsos,
  M.}
\newblock Is {P2P} dying or just hiding?
\newblock In {\em Proceedings of the GLOBECOM 2004 Conference\/} (Dallas,
  Texas, November 2004), IEEE Computer Society Press.

\bibitem{Lee:2003}
{\sc Lee, C.~B., Roedel, C., and Silenok, E.}
\newblock Detection and characterization of port scan attacks, 2003.

\bibitem{Liu:2009}
{\sc Liu, Y., and Pan, J.}
\newblock The impact of {NAT} on {B}it{T}orrent-like {P2P} systems.
\newblock In {\em Proceedings of 2009 9th IEEE International Conference on
  Peer-to-Peer Computing\/} (2009), pp.~242--251.

\bibitem{Northcutt:2002}
{\sc Northcutt, S., and Novak, J.}
\newblock {\em Network Intrusion Detection: An Analyst's Handbook}, 3rd~ed.
\newblock New Riders Publishing, Thousand Oaks, CA, USA, 2002.

\bibitem{opaque:2009}
{\sc opaque}.
\newblock Internet study 2008/09, 2009.

\bibitem{Paxson:1999}
{\sc Paxson, V.}
\newblock Bro: A system for detecting network intruders in real-time.
\newblock {\em Computer Networks 31\/} (December 1999), 2435--2463.

\bibitem{Roesch:1999}
{\sc Roesch, M.}
\newblock Snort: Lightweight intrusion detection for networks.
\newblock In {\em Proceedings of LISA '99: 13th Systems Administration
  Conference\/} (1999), USENIX, pp.~229--238.

\bibitem{Wu:2010}
{\sc Wu, D., Dhungel, P., Hei, X., Zhang, C., and Ross, K.~W.}
\newblock Understanding peer exchange in {B}it{T}orrent systems.
\newblock In {\em Proceedings of 2010 10th IEEE International Conference on
  Peer-to-Peer Computing\/} (2010), pp.~1--8.

\end{thebibliography}

\appendix

\section{Bro Modifications}

\begin{lstlisting}[caption=\texttt{raw\_ns\_bytes\_to\_uint16} helper function for Bro,
                 label=lst:bro_raw_ns_bytes_to_uint16]
function raw_ns_bytes_to_uint16%(b: string%): count
	%{
	uint32 u = 0;

	if ( b->Len() < 2 )
		builtin_error("too short a string as input to raw_bytes_to_uint16()");

	else
		{
		const u_char* bp = b->Bytes();
		u = (bp[0] << 8) | bp[1];
		}

	return new Val(u, TYPE_COUNT);
	%}
\end{lstlisting}

\begin{lstlisting}[caption=\texttt{raw\_nl\_bytes\_to\_uint32} helper function for Bro,
                 label=lst:bro_raw_nl_bytes_to_uint32]
function raw_nl_bytes_to_uint32%(b: string%): count
	%{
	uint32 u = 0;

	if ( b->Len() < 4 )
		builtin_error("too short a string as input to raw_bytes_to_uint32()");

	else
		{
		const u_char* bp = b->Bytes();
		u = (bp[0] << 24) | (bp[1] << 16) | (bp[2] << 8) | bp[3];
		}

	return new Val(u, TYPE_COUNT);
	%}
\end{lstlisting}

\begin{lstlisting}[caption=\texttt{sub\_bytes\_sane} helper function for Bro,
                 label=lst:bro_sub_bytes_sane]
function sub_bytes_sane%(s: string, start: count, n: int%): string
	%{
	BroString* ss = s->AsString()->GetSubstring(start, n);

	if ( ! ss )
		ss = new BroString("");

	return new StringVal(ss);
	%}
\end{lstlisting}

\begin{lstlisting}[caption=\texttt{raw\_byte} helper function for Bro,
                 label=lst:bro_raw_byte]
function raw_byte%(s: string, index: count%): count
	%{
	return new Val(s->Bytes()[index], TYPE_COUNT);
	%}
\end{lstlisting}

\begin{lstlisting}[caption=\texttt{raw\_bytes} helper function for Bro,
                 label=lst:bro_raw_bytes]
function raw_bytes%(s: string%): index_vec
	%{
	VectorVal* result_v =
		new VectorVal(new VectorType(base_type(TYPE_COUNT)));

	int i;
	for ( i = 0; i < s->Len(); ++i )
		{
		result_v->Assign(i, new Val(s->Bytes()[i], TYPE_COUNT), 0);
		}

	return result_v;
	%}
\end{lstlisting}

\subsection{HTTP Tracker}

\begin{lstlisting}[caption=\texttt{tcp\_contents} event handler for HTTP analyzer,
                 label=lst:httptracker_tcp_contents,
                 basicstyle=\tiny]
event tcp_contents(c: connection, is_orig: bool, seq: count, contents: string)
	{
	for ( s in find_all(contents, /5\:peers[0-9]+\:/) )
		{
		local sv: string_vec = str_split(s, vector(7));
		local len: count = to_count(sub_bytes_sane(sv[2], 0, byte_len(sv[2]) - 1));

		local i: count = 0;
		local peers_compact: string = sub_bytes_sane(contents, strstr(contents, s) + byte_len(s) - 1, len);
		for ( dummy in peers_compact )
			{
			if ( i % 6 == 0 )
				{
				local peer_compact: string = sub_bytes_sane(peers_compact, i - 6, 6);
				local peer: TrackerPeer;

				peer$h = raw_bytes_to_v4_addr(peer_compact);
				peer$p = raw_ns_bytes_to_uint16(sub_bytes_sane(peer_compact, 4, 2));

				local orig: addr = ( is_orig ? c$id$resp_h : c$id$orig_h );

				if ( orig !in peer_mappings )
					{
					peer_mappings[orig] = set();
					}
				add peer_mappings[orig][peer];
				}

			i = i + 1;
			}
		}
	}
\end{lstlisting}

\subsection{UDP Tracker}

\subsubsection{Packet structure}

The following information was retrieved from BitTorrent Enhancement Proposal 15
  \footnote{\url{http://bittorrent.org/beps/bep_0015.html}}.

\begin{table}[t]
  \centering
  {\tiny
  \begin{tabular}{ | c | c | c | }
    \hline
    \textbf{Offset} & \textbf{Size} & \textbf{Name} \\ \hline
    \hline
    $0$                & 32-bit integer  & \texttt{action} \\ \hline
    $4$                & 32-bit integer  & \texttt{transaction\_id} \\ \hline
    $8$                & 32-bit integer  & \texttt{interval} \\ \hline
    $12$               & 32-bit integer  & \texttt{leechers} \\ \hline
    $16$               & 32-bit integer  & \texttt{seeders} \\ \hline
    $20 + 6 \times n$  & 32-bit integer  & IP address \\ \hline
    $24 + 6 \times n$  & 16-bit integer  & TCP port \\
    \hline
  \end{tabular}
  }
  \caption{Packet structure for a UDP tracker \texttt{announce\_response}
           containing $n$ peers}
  \label{tab:udptracker-announce-response}
\end{table}

\subsubsection{Code}

\begin{lstlisting}[caption=\texttt{udp\_contents} event handler for UDP Tracker analyzer,
                 label=lst:udptracker_udp_contents,
                 basicstyle=\tiny]
event udp_contents(u: connection, is_orig: bool, contents: string)
	{
	local len: count = byte_len(contents);

	if ( ( len - 20 ) % 6 != 0 )
		{
		return;
		}

	if ( sub_bytes_sane(contents, 0, 4) != "\x00\x00\x00\01" )
		{
		return;
		}

	announce_response(u, is_orig, contents);
	}
\end{lstlisting}

\subsection{Azureus DHT}

\subsubsection{Packet structures}

The following information was retrieved from the Vuze Wiki
  \footnote{\url{http://wiki.vuze.com/w/Distributed_hash_table}}.

\begin{table*}[t]
  \centering
  {\tiny
  \begin{tabularx}{\textwidth}{ | c | c | X | }
    \hline
    \textbf{Name} & \textbf{Width} & \textbf{Note} \\ \hline
    \hline
    \texttt{byte} & 1 byte & Single byte \\ \hline
    \texttt{short} & 2 bytes & Big endian \\ \hline
    \texttt{int} & 4 bytes & Big endian \\ \hline
    \texttt{long} & 8 bytes & Big endian \\ \hline
    \texttt{boolean} & 1 byte & False = 0; True = 1 \\ \hline
    \texttt{address} & 7 bytes or 19 bytes & First byte indicates length of the IP address (4 for IPv4, 16 for IPv6); next comes the address in network byte order; the last value is port number as short \\ \hline
    \texttt{contact} & 9 bytes or 21 bytes & First byte indicates contact type, which must be UDP (1); second byte indicates the contact's protocol version; the rest is an address \\
    \hline
  \end{tabularx}
  }
  \caption{Serialization for ADHT values}
  \label{tab:adht-serialization}
\end{table*}

\begin{table*}[t]
  \centering
  {\tiny
  \begin{tabularx}{\textwidth}{ | c | c | c | X | }
    \hline
    \textbf{Name} & \textbf{Type} & \textbf{Protocol Version} & \textbf{Note} \\ \hline
    \hline
    \texttt{CONNECTION\_ID} & \texttt{long} & always & Random number with most significant bit set to 1 \\ \hline
    \texttt{ACTION} & \texttt{int} & always  & Type of the packet \\ \hline
    \texttt{TRANSACTION\_ID} & \texttt{int} & always & Unique number used through the communication; it is randomly generated at the start of the application and increased by 1 with each sent packet \\ \hline
    \texttt{PROTOCOL\_VERSION} & \texttt{byte} & always & Version of protocol used in this packet \\ \hline
    \texttt{VENDOR\_ID} & \texttt{byte} & $\geq$\texttt{VENDOR\_ID} & ID of the DHT implementator; 0 = Azureus, 1 = ShareNet, 255 = unknown \\ \hline
    \texttt{NETWORK\_ID} & \texttt{int} & $\geq$\texttt{NETWORKS} & ID of the network; 0 = stable version; 1 = CVS version \\ \hline
    \texttt{LOCAL\_PROTOCOL\_VERSION} & \texttt{byte} &
    $\geq$\texttt{FIX\_ORIGINATOR} & Maximum protocol version this node supports; if this packet's protocol version is $<$FIX\_ORIGINATOR then the value is stored at the end of the packet \\ \hline
    \texttt{NODE\_ADDRESS} & \texttt{address} & always & Address of the local node \\ \hline
    \texttt{INSTANCE\_ID} & \texttt{int} & always & Application's helper number; randomly generated at the start \\ \hline
    \texttt{TIME} & \texttt{long} & always & Time of the local node; stored as number of milliseconds since Epoch \\
    \hline
  \end{tabularx}
  }
  \caption{Header for ADHT requests}
  \label{tab:adht-request-header}
\end{table*}

\begin{table*}[t]
  \centering
  {\tiny
  \begin{tabularx}{\textwidth}{ | c | c | c | X | }
    \hline
    \textbf{Name} & \textbf{Type} & \textbf{Protocol Version} & \textbf{Note} \\ \hline
    \hline
    \texttt{ACTION} & \texttt{int} & always  & Type of the packet \\ \hline
    \texttt{TRANSACTION\_ID} & \texttt{int} & always  & Must be equal to \texttt{TRANSACTION\_ID} from the request \\ \hline
    \texttt{CONNECTION\_ID} & \texttt{long} & always  & Must be equal to \texttt{CONNECTION\_ID} from the request \\ \hline
    \texttt{PROTOCOL\_VERSION} & \texttt{byte} & always  & Version of protocol used in this packet \\ \hline
    \texttt{VENDOR\_ID} & \texttt{byte} & $\geq$\texttt{VENDOR\_ID}  & Same meaning as in the request \\ \hline
    \texttt{NETWORK\_ID} & \texttt{int} & $\geq$\texttt{NETWORKS}  & Same meaning as in the request \\ \hline
    \texttt{INSTANCE\_ID} & \texttt{int} & always  & Instance id of the node that replies to the request \\
    \hline
  \end{tabularx}
  }
  \caption{Header for ADHT replies}
  \label{tab:adht-reply-header}
\end{table*}

\subsubsection{Code}

\begin{lstlisting}[caption=\texttt{udp\_contents} event handler for ADHT analyzer,
                 label=lst:adht_udp_contents,
                 basicstyle=\tiny]
event udp_contents(u: connection, is_orig: bool, contents: string)
	{
	local protocol_version: count;
	local action: string;

	protocol_version = raw_byte(contents, 16);
	if ( protocol_version in valid_protocol_versions )
		{
		local offset: count = 17;
		if ( protocol_version >= protocol_versions["VENDOR_ID"] )
			{
			offset += 1;
			}
		if ( protocol_version >= protocol_versions["NETWORKS"] )
			{
			offset += 1;
			}
		if ( protocol_version >= protocol_versions["FIX_ORIGINATOR"] )
			{
			offset += 4;
			}

		offset += 5;

		local p: count;
		p = raw_ns_bytes_to_uint16(sub_bytes_sane(contents, offset, 2));
		if ( p == port_to_count(u$id$orig_p) )
			{
			action = sub_bytes_sane(contents, 8, 4);

			if ( action == "\x00\x00\x04\x04" )
				{
				find_node_request(u, is_orig, contents);
				}
			if ( action == "\x00\x00\x04\x06" )
				{
				find_value_request(u, is_orig, contents);
				}
			}
		}

	if ( sub_bytes_sane(contents, 4, 4) in transaction_ids )
		{
		local a: DhtAction = transaction_ids[sub_bytes_sane(contents, 4, 4)];

		action = sub_bytes_sane(contents, 0, 4);

		if ( a == FindNode && action == "\x00\x00\x04\x05" )
			{
			find_node_response(u, is_orig, contents);
			}
		if ( a == FindValue && action == "\x00\x00\x04\x07" )
			{
			find_value_response(u, is_orig, contents);
			}
		}
	}
\end{lstlisting}

\subsection{Mainline DHT}

\begin{lstlisting}[caption=\texttt{mdht\_find\_values} helper function for MDHT analyzer,
                 label=lst:mdht_find_values,
                 basicstyle=\tiny]
function parse_values(u: connection, is_orig: bool, contents: string)
	{
	for ( s in find_all(contents, /6\:valuesl6\:/) )
		{
		local offset: count = strstr(contents, s) + 8;

		for ( c in contents )
			{
			if ( sub_bytes_sane(contents, offset, 2) != "6:" )
				{
				break;
				}
			offset += 2;

			local peer_compact: string = sub_bytes_sane(contents, offset, 6);
			local peer: DhtPeer;

			peer$h = raw_bytes_to_v4_addr(peer_compact);
			peer$p = raw_ns_bytes_to_uint16(sub_bytes_sane(peer_compact, 4, 2));

			if ( u$id$orig_h !in peer_mappings )
				{
				peer_mappings[u$id$orig_h] = set();
				}
			add peer_mappings[u$id$orig_h][peer];

			offset += 6;
			if ( offset > byte_len(contents) )
				{
				break;
				}
			}
		}
	}
\end{lstlisting}

\subsection{Other DHTs}

\begin{lstlisting}[caption=\texttt{udp\_contents} helper function for BTUDP analyzer,
                 label=lst:btudp_udp_contents,
                 basicstyle=\tiny]
event udp_contents(u: connection, is_orig: bool, contents: string)
	{
	local matched: bool = F;
	local msb4: string = sub_bytes_sane(contents, 0, 4);
	local len: count = byte_len(contents);

	if ( msb4 == /d1:[areqt]/ )
		{
		matched = T;
		}

	if ( msb4 == /d1.:/ )
		{
		matched = T;
		}

	# other checks for signature matching ommitted for brevity ...

	if ( !matched )
		{
		return;
		}

	local peer: BTUDPPeer;
	local orig: addr;

	if ( is_orig )
		{
		orig = u$id$orig_h;
		peer$h = u$id$resp_h;
		peer$p = port_to_count(u$id$resp_p);
		}
	else
		{
		orig = u$id$resp_h;
		peer$h = u$id$orig_h;
		peer$p = port_to_count(u$id$orig_p);
		}

	if ( orig !in peer_mappings )
		{
		peer_mappings[orig] = set();
		}
	add peer_mappings[orig][peer];
	}

\end{lstlisting}

\subsection{UTPEX}

\begin{lstlisting}[caption=\texttt{tcp\_contents} event handler for UTPEX analyzer,
                 label=lst:utpex_tcp_contents,
                 basicstyle=\tiny]
event tcp_contents(c: connection, is_orig: bool, seq: count, contents: string)
{
	for ( s in find_all(contents, /5\:added[0-9]+\:/) )
		{
		local sv: string_vec = str_split(s, vector(7));
		local len: count = to_count(sub_bytes_sane(sv[2], 0, byte_len(sv[2]) - 1));

		local i: count = 0;
		local peers_compact: string = sub_bytes_sane(contents, strstr(contents, s) + byte_len(s) - 1, len);
		for ( dummy in peers_compact )
			{
			if ( i % 6 == 0 )
				{
				local peer_compact: string = sub_bytes_sane(peers_compact, i - 6, 6);
				local peer: PexPeer;

				peer$h = raw_bytes_to_v4_addr(peer_compact);
				peer$p = raw_ns_bytes_to_uint16(sub_bytes_sane(peer_compact, 4, 2));

				local orig: addr = ( is_orig ? c$id$resp_h : c$id$orig_h );

				if ( orig !in peer_mappings )
					{
					peer_mappings[orig] = set();
					}
				add peer_mappings[orig][peer];
				}

			i = i + 1;
			}
		}
}
\end{lstlisting}

\subsection{Port/Peer Ratio}

\begin{lstlisting}[caption=\texttt{check\_scan} alarm suppression for PPR,
                 label=lst:btscan_check_scan,
                 basicstyle=\tiny]
function check_scan(c: connection, established: bool, reverse: bool): bool
	{

	# lines omitted for brevity ...
	
	local n = |distinct_peers[orig]|;
	local p = orig in distinct_ports ? |distinct_ports[orig]| * 1.0 : 1.0;
	local pn = p / n;

	# Check for threshold if not outbound.
	if ( ! shut_down_thresh_reached[orig] &&
     		n >= shut_down_thresh &&
     		! outbound && orig !in Site::neighbor_nets &&
     		(pn < port_peer_ratio_thresh_lower || pn > port_peer_ratio_thresh_upper) )
		{
		shut_down_thresh_reached[orig] = T;
		local msg = fmt("shutdown threshold reached for %s", orig);
		NOTICE([$note=ShutdownThresh, $src=orig,
			$identifier=fmt("%s", orig),
			$p=service, $msg=msg]);
		}
	
	# lines omitted for brevity ...

	}
\end{lstlisting}

\end{document}